\documentclass[pra,superscriptaddress,showpacs,twocolumn]{revtex4}

\usepackage{amsmath}
\usepackage{graphicx}
\usepackage{epstopdf}
\usepackage{amsfonts} 
\usepackage{hyperref}
\usepackage{bm}

\newcommand{\pd}[2]{\frac{\partial #1}{\partial #2}}

\newcommand{\GF}[2]{G_{{#1}{#2}}(\bm{r}, \bm{r}', \omega)}
\newcommand{\fGF}[2]{\tilde{G}_{{#1}{#2}}(\bm{k}_p, \omega; z, z')}
\newcommand{\fGFZ}[2]{\tilde{G}_{{#1}{#2}}}

\begin{document}

\title{Superradiance, subradiance, and suppressed superradiance\\ of dipoles near a metal interface}

\author{J. J. Choquette}
\affiliation{Institute for Quantum Information Science,
        University of Calgary, Calgary, Alberta T2N 1N4, Canada}
\affiliation{Department of Physics, St. Francis Xavier University,
  Antigonish, Nova Scotia B2G 2W5, Canada} 

\author{Karl-Peter Marzlin}
\affiliation{Department of Physics, St. Francis Xavier University,
  Antigonish, Nova Scotia B2G 2W5, Canada}
\affiliation{Institute for Quantum Information Science,
        University of Calgary, Calgary, Alberta T2N 1N4, Canada}

\author{B. C. Sanders}
\affiliation{Institute for Quantum Information Science,
        University of Calgary, Calgary, Alberta T2N 1N4, Canada}

\begin{abstract}
We theoretically characterize the collective radiative behaviour of $N$ classical emitters near an interface 
between different dielectrics that supports the transfer of surface plasmon modes 
into the far-field of electromagnetic radiation. The phenomena of
superradiance and surface plasmons can be combined to amplify the emitted
radiation intensity $S$ as $S= A N^2 S_0$ compared to a single emitter's intensity $S_0$ in free space. For a practical case study within the paper $A=240$, compared to $A=1$ in free space.
We furthermore demonstrate
that there are collective modes 
for which the intensity of the emitted radiation is suppressed by two orders of magnitude despite
their supperadiant emission characteristics.
A method to control the emission characteristics of the system and to switch from super- to sub-radiant
behaviour with a suitably detuned external driving field is devised.
\end{abstract}


\maketitle

\section{Introduction}

Observing light spontaneously emitted from a system of atoms or molecules is a versatile tool to determine the system's dynamics and has a number of important applications. In atomic physics it can generally be used to learn about the state of atoms and molecules \cite{CohenTannBook}. In quantum information it helps to detect entanglement between ions or atoms \cite{ionTrapCitation, entangledAtomicVaporsCitation?}. In biophysics it is used to identify and track chemical compounds in biological tissues. 
While spontaneous emission is well understood for isolated emitters, emission from a system of emitters
varies extensively with their interaction, depending on the state of nearby emitters and the surrounding media. 
Only by understanding the role of interactions in light emission can the state reliably be determined. The process of the collective decay rate being coherently increased is known as superradiance \cite{Dicke}, and has received renewed interest with the development of nanotechnologies \cite{collQuantumEffects}. In quantum optics, superradiance can be used to create entanglement between atoms and light \cite{SREntanglement}. Entanglement within a superradiant system is also being used to understand phase transitions in quantum systems \cite{SRQuantumPhase}. For this reason there is strong interest in methods that affect the interactions in cooperative emission.

A simple method to influencing the interaction strength is to place the system near a metal interface that supports surface plasmon modes \cite{morowitz}. 
Surface plasmons are
charge density waves confined to a very small region close to the
interface. The field of surface plasmons are consequently very
large.  This effect has
long been used in techniques such as surface enhanced Raman spectroscopy
\cite{Fleischman,JLD}. In quantum
information science, surface plasmon modes on a nanowire 
are proposed to create
single-photon sources and transistors \cite{ML1,ML2} and
enhance the emission properties of light emitting diodes
\cite{Vuckovic}. Furthermore, the coupling of emitters to surface plasmon modes allows the collective excitation of superradiant surface plasmons \cite{Bonifacio}.

\begin{figure}
\begin{center}
\includegraphics[width=3.5cm]{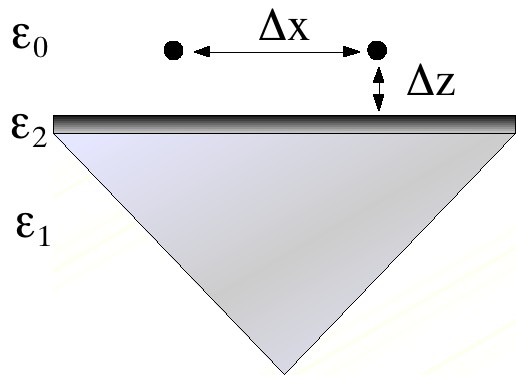}
\caption{\label{fig:interface}
  Kretschmann configuration of a thin metal film
  interface between a prism and vacuum used to couple emitters to
  surface plasmon modes. Region 1 consists of a prism
  ($\sqrt{\varepsilon_1}=1.51$), region 2 corresponds to a thin metal
  film ($\sqrt{\varepsilon_2}=0.08 + 4.8 i$), and region 3 is the
  vacuum.} 
\end{center}
\end{figure}

In this paper we show how superradiance and surface plasmons can be combined to increase the interaction strength between
matter and light by several orders of magnitude. We consider emitters near a planar metal interface in the Kretschmann configuration for attenuated total reflection \cite{Kretschmann},
see Fig.~\ref{fig:interface}. 
In this situation, surface plasmons induce a greater coupling of radiation into the far-field
for a certain direction (the surface plamon resonance angle) where the flux of radiation is two orders of magnitude larger \cite{IntEnh}. 
Furthermore, the increased radiation energy density in the near field of surface plamons leads to a stronger coupling
between different emitters \cite{SRPlasmonsNano} so that superradiance is further increased. We characterize the collective
radiative eigenmodes of a set of oscillating emitters and demonstrate how their emission can be manipulated from super- to
a sub-radiant behaviour.

After a brief review of superradiance and surface plasmons in Sec.~\ref{sec:backg} 
and our model assumptions in Sec.~\ref{sec:model} we will consider 
harmonically oscillating emitters that are driven by a light field in Sec.~\ref{sec:driven}.
We demonstrate that their collective dynamics can be changed from super- to
sub-radiant behaviour by changing the frequency of the field. In Sec.~\ref{sec:decay}
we study the collective decay of initially excited harmonic oscillators and
give a detailed description of super- and sub-radiant decay modes, their
decay rates, emission pattern, and discuss how well superradiant emission
is realized in these modes. Three appendices contain details of our derivations.

\section{Background}\label{sec:backg}

Superradiance is a cooperative effect in which the radiation 
emitted by $N$ scatterers is substantially enhanced
compared to isolated scatterers \cite{Dicke}.
For this effect to occur the distance between the emitters needs to be
much smaller than the wavelength of the radiation. The field amplitudes
of the emitters then interfere constructively so that the radiation intensity
is increased by a factor of $N^2$ compared 
to a single emitter. Furthermore, light emitted by one of the scatterers
can induce stimulated emission for another scatterer so that the collective
decay rate is enhanced by a factor of $N$ if all emitters are initially excited.
These two defining properties of superradiant emission are consequences of
phasing and radiation reaction which can occur both in classical
and in quantum systems \cite{LJM}. 
Radiation reaction is generally affected by any surrounding dielectric material.
In particular, when atom-sized dipolar emitters are placed within a wavelength
of a metal interface, surface plasmons can be generated.

Surface plasmons consist of electromagnetic fields that accompany 
electron density oscillations (the plasmon) on the surface of a metal. Because the surface plasmon is confined within a nm-sized region on the interface the electromagnetic field is concentrated. This concentration depends strongly on the geometry and the dielectric properties of the interface. Because of the increased intensity of surface plasmon fields, an emitter reacts strongly with surface plasmon modes when placed near a metal film and the radiative properties of the emitter are significantly affected\cite{raether}.
In the following we will study in detail how surface plasmons affect superradiance
of emitters near a planar interface.

\section{Oscillating emitters near a metal interface}\label{sec:model}
We consider the behavior of $N$ atomic dipoles that are located
near a thin metallic interface that supports 
surface plasmon modes, see Fig.~\ref{fig:interface}. 
To simplify the considerations we study an infinite planar interface so that
the prism, which is needed in experiments for phase-matching, can be
replaced by an infinite half-space filled with a non-absorbing dielectric.
The dipoles are
located at positions $\bm{r}^{(1)},...,\bm{r}^{(N)}$
in region 3 above the interface. 

To describe the system of emitters we model each dipole as a
charged classical  
harmonic oscillator with charge $q$, mass $m$, and  resonance frequency of $\omega_0$. The restoring force associated with the
harmonic motion originates from the attraction by a stationary charge distribution of opposite sign 
(the ``nucleus'') located at the point $\bm{r}^{(i)}$.
Charge $i$ oscillates about the
point $\bm{r}^{(i)}$ with an amplitude $\bm{x}^{(i)}$ 
that is of the order of Bohr's radius and much smaller than the individual separation distance between emitters $\Delta x$.
The equation of motion for the harmonic oscillators affected by an
external electric field  $\bm{E}(\bm{r},t)$ 
is derived in Appendix \ref{app:fieldEqn},
\begin{equation}
\label{eqn:coupDE}
  \ddot{\bm{x}}_{0}^{(j)}+\omega_0^2\bm{x}_{0}^{(j)} = 
  \frac{q}{m} \left (
  \bm{E}(\bm{r}^{(j)},t) +
  \sum_{i=1}^N \bm{E}^{(i)}(\bm{r}^{(j)},t)
  \right ).
\end{equation}
The term in parentheses
can be interpreted as the local
electric field at the position $\bm{r}^{(j)}$ of the $j$th oscillator. 
It consists of the external applied field $ \bm{E}(\bm{r}^{(j)},t)$
in region 3 (e.g., a laser beam) plus the superposition of the fields 
$ \bm{E}^{(i)}(\bm{r}^{(j)},t)$ that are created by emitter $i$. 
Due to the presence of
the interface the latter part  of the contribution is more involved and is 
conveniently expressed by
\begin{equation} 
  \bm{E}^{(j)}(\bm{r}, t) =   
  -\mu_0q  \int_{-\infty}^{\infty}dt' 
  \bm{G}(\bm{r}, \bm{r}^{(j)},t-t' )\,
  \ddot{\bm{x}}^{(j)}(t'),
\label{EfieldJ}\end{equation}

The dyadic retarded Green's function $ \bm{G}$ describes the propagation of electromagnetic fields
in the presence of the interface and takes into account the
corresponding boundary conditions.
It is frequently used to describe radiative systems with boundary conditions  \cite{CHOGreenF,PRA53:1818,gan:optExpr2006} 
despite that fact that its explicit form is rather involved  (see Appendix \ref{sec:greenf}).
This is because the Green's function contains all information about the propagation of electromagnetic waves between 
two arbitrary points in the presence of the interface, which is necessary to analyze emitters at different positions. 
Alternatively one could use an expansion of 
the electromagnetic field in terms of radiative field modes \cite{PhysRevA.68.023809} which has been shown
to be equivalent \cite{DKW}.

Within the Green's function formalism, the effect of surface plasmons are included through the Fresnel coefficients, i.e., the complex
ratio of the reflected ($R^\text{TM}_{i,i-1}$) or transmitted ($T^\text{TM}_{i,i-1}$) electric field and the incident field, where $i$ is region in consideration and $i-1$ the neighbouring region.
Here ``TM'' refers to transverse magnetic polarization which is necessary to generate surface plasmons.
The Fresnel coefficients appear explicitly in the Green's function
(see  Appendix \ref{sec:greenf}); surface plasmons generate characteristic resonances in these coefficients.

The dispersion relation for a surface plasmon at a single vacuum-metal interface,
 \begin{equation}
 \label{eqn:spk}
 c^2 k_\text{sp}^2=\left(\frac{\omega}{c}\right)^2\frac{\varepsilon_0 \varepsilon_2}{\varepsilon_0+\varepsilon_2},
 \end{equation}
can be derived from the condition that there is no reflected field, $R^\text{TM}_{3,2}=0$ \cite{raether}.
Here $k_\text{sp}$ is the surface plasmon wave vector component parallel to the interface.
For a single interface between only two different dielectric media the dispersion relation
corresponds to the poles of the Green's function in frequency space.
Even though the inclusion of a second interface changes the dispersion relation, Eq.~(\ref{eqn:spk}) is often presented within the context of both single and double interfaces \cite{NegmTalaat}. 
This approximation suffices to determine the surface plasmon resonance angle in 
attenuated total reflection for thin films. However, it omits complex valued contributions to the wavevector component $k_\text{sp}$\cite{raether,Davis},
most notably complex contributions that would correspond to radiative coupling to the prism (i.e., the third medium needed in
attenuated total reflection).  
For the Kretschmann configuration the narrow resonance angle of surface plasmons results 
in an increased radiation density which can be used to increase the coupling between
emitters and radiation by several orders of magnitude. It is the combination of this effect
with super-radiance that is at the centre of our attention.

\section{Super- and sub-radiant emission by driven oscillators}\label{sec:driven}
If the emitters are driven by an external electric field
of the form $\bm{E}(\bm{r},t) =\bm{E}_0e^{i\bm{k}\cdot \bm{r}}e^{-i\omega{t}}$
they will eventually settle into a state where each dipole oscillates
with the frequency $\omega$ of the driving field, so that
$\bm{x}(t)=\bm{x}_0e^{-i\omega t}$
\footnote{To simplify the derivations we use a complex amplitude $\bm{x}(t) $ for the oscillators. The use of complex solutions is possible because
for harmonic oscillators interacting via the electromagnetic field the equations of motions are linear second-order differential
equations. Real and imaginary part of $\bm{x}(t) $ then correspond to two linearly independent, real solutions of the
equations of motion.
For the field intensity we use the quantity $| \bm{E} |^2$ which is proportional to the intensity for real
oscillators amplitudes averaged over one cycle $2\pi /\omega$.
}.
It is shown
in Appendix \ref{app:fieldEqn} that this ansatz transforms
Eq.~(\ref{eqn:coupDE}) into the algebraic equation
\footnote{
 Eq.~(\ref{eqn:n3osc}) corresponds to Eq.~ (\ref{algEq1})
in the limit of stationary amplitudes, i.e., 
$\Gamma = 0$ in Eq.~(\ref{algEq1}).
}
\begin{equation} 
\label{eqn:n3osc}
\begin{split}
   \bm{E}_0^{(i)} =&
  e^{-i\bm{k}\cdot\bm{r}}\Bigg[ \frac{m}{q} (\omega_0^2-\omega^2) \delta_{ij}
  \openone-
 \mathrm{Im}{\bm{G}}(\bm{r}^{(i)},
{\bm{r}^{(i)}}, \omega)\\
 & \times  
 \delta_{ij} \frac{\mu_0\omega^2q}{2\pi}
  - \frac{\mu_0\omega^2q}{2\pi}
  \sum_{j\neq{i}}^N{\bm{G}}(\bm{r}^{(i)},
  \bm{r}^{(j)}, \omega) \Bigg ]\bm{x}^{(j)}_0,
\end{split} 
\end{equation}  
and forms a set of $3N$ coupled equations for the Cartesian
components
of $N$ oscillators. The indices $i,j$ run from 1 to $N$ and
$\openone$ denotes the unit matrix
in three dimensions. To find the total electric field 
emitted by all
$N$ sources we need to solve these equations for
$\bm{x}^{(i)}$ and then superpose the fields produced by 
all sources.

\begin{figure}
\begin{center}
\centerline{{\bf \hspace{0.4cm}(a)\hspace{3.5cm} (b)}}
\includegraphics[width=4.2cm]{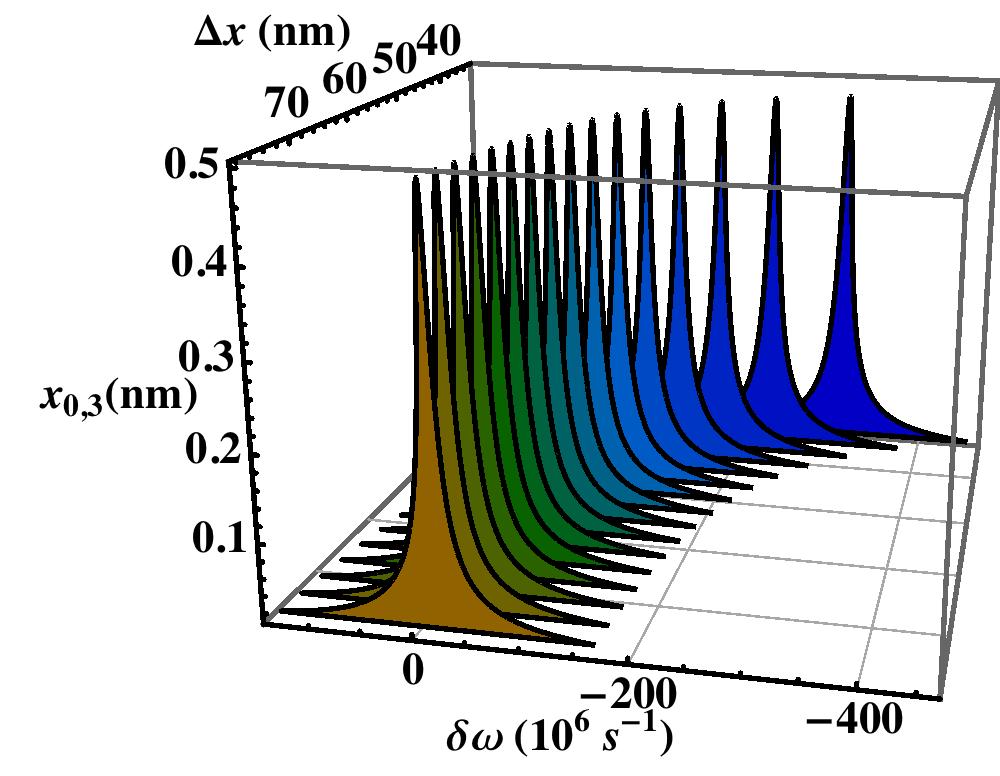}
\includegraphics[width=4.2cm]{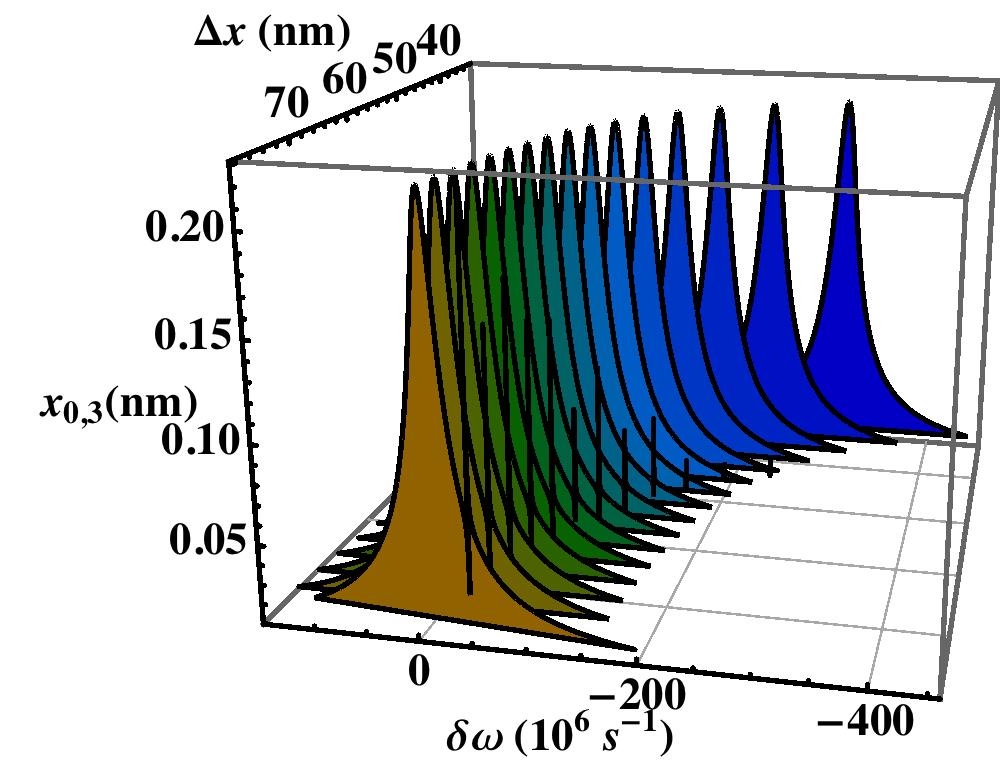}
\caption{\label{fig:ampresp}
The amplitude response in $z$-direction of two emitters in (a)
free space (b) and  placed at a distance
of $\Delta z=100$nm from the metal interface and 
separated a distance $\Delta x$ from each other.
Both emitters are driven in the $z$-direction by an electric field 
that is detuned by $\delta \omega$ from a resonance frequency of $\omega_0=2.36\times 10^{15}s^{-1}$.}
\end{center}
\end{figure}

Figure \ref{fig:ampresp}(a) shows the amplitude response $x_{0,3}$,
i.e., the $z$-component of the amplitude vector  
$\bm{x}_0 $ of the oscillators, for
two nearby emitters in free space driven by an electric field
polarized in the $z$ direction. 
For separation distances $\Delta x$ less than $50$ nm the resonance frequency begins
to shift considerably due to dynamic dipole-dipole coupling between the emitters.
Because in free space the dyadic Green's function is diagonal with respect to radiation polarization,
and because the emitters are only driven along the z-axis, 
there is only coupling between the $z$-components of oscillations. 

\begin{figure}
\begin{center}
\centerline{{\bf \hspace{0.4cm}(a)\hspace{3.5cm} (b)}}
\includegraphics[width=8.5cm]{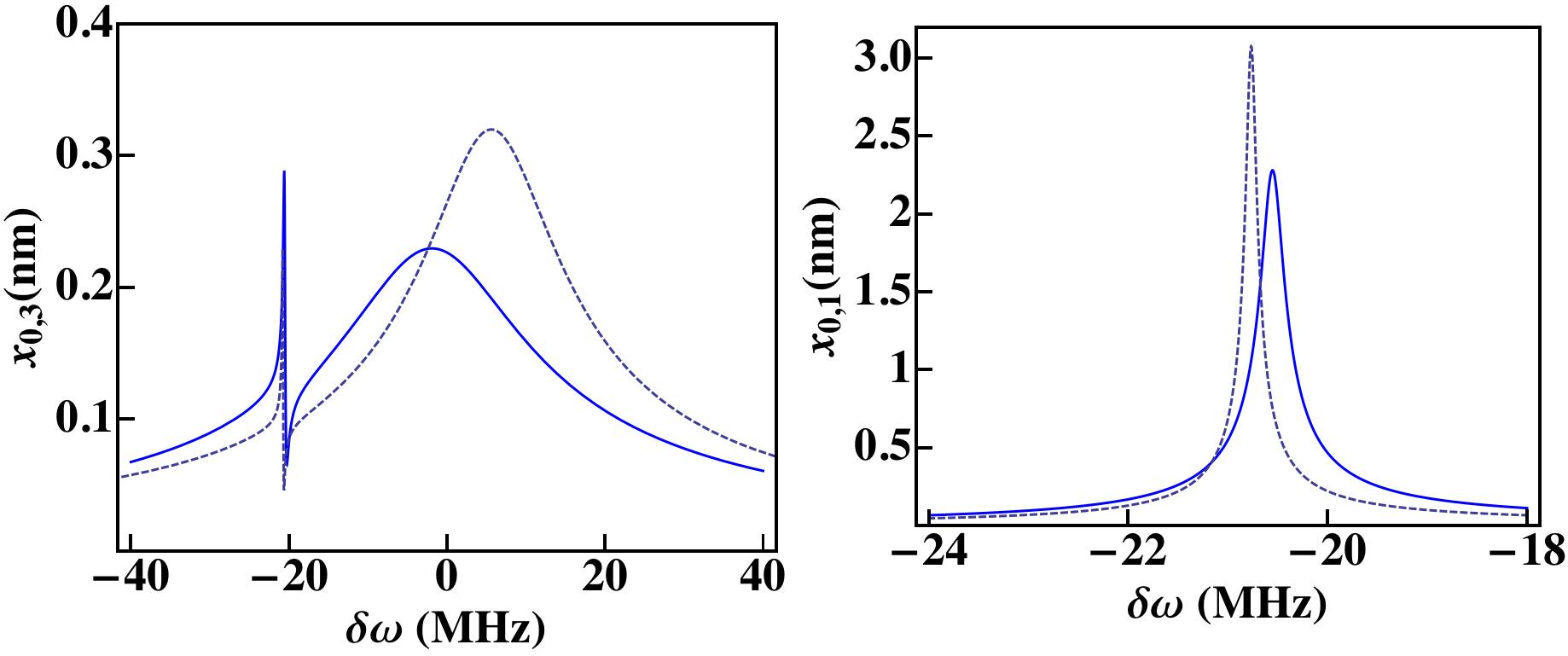}
\caption{\label{fig:mirintresp}The $z$-amplitude $x_{0,3}$ (a) and $x$-amplitude $x_{0,1}$ (b) of two emitters near a metal film (solid line)
and a perfect mirror (dashed line). The emitters are driven by a field that is detuned by $\delta\omega$.
They are separated by $\Delta x=80$nm from each other and $\Delta z=100$nm from the metal interface.}
\end{center}
\end{figure}

Placing the two emitters close to the metal interface leads to a dramatic change
in the emitter dynamics.
In Fig.~\ref{fig:ampresp}(b) the $z$-amplitude of two scatterers that are a distance $\Delta z = 100$nm 
away from the interface is displayed. It has a similar overall shape as the
amplitude response in free space, but an
additional narrow resonance peak appears on the tail of the primary
resonance. This narrow feature is more prominent in Fig. \ref{fig:mirintresp}(a)
which displays the same quantity
for a fixed distance  $\Delta x = 80$nm between the scatterers.

The origin of the secondary resonance is the reflection of light emitted by
one oscillator from the interface and its subsequent absorption
by the other oscillator. In free space light emitted by one oscillator necessarily
has the same polarization as the external driving field. Therefore, the emitters
are collectively oscillating along the $z$-direction. However, light
that is reflected by the interface can have a different polarization.
In the case under consideration it induces a coupling between the
$x$- and $z$-components of the oscillators; mathematically
this is related to the $G_{13}$ terms in Eq.~(\ref{eqn:n3osc}). 

\begin{figure}
\begin{center}
\centerline{{\bf \hspace{0.7cm}(a)\hspace{3.5cm} (b)}}\vspace{2mm}
\includegraphics[width=4cm]{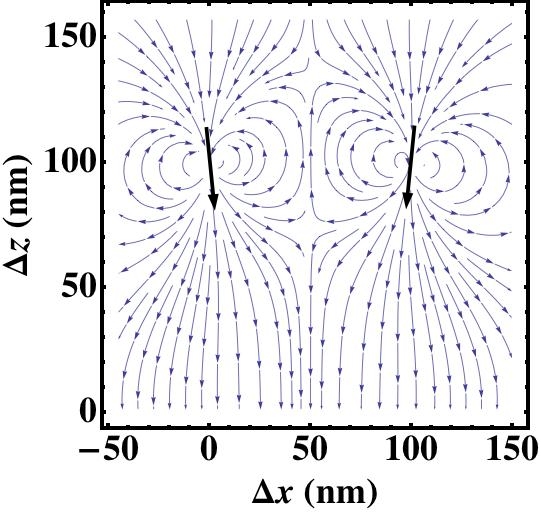}
\includegraphics[width=4cm]{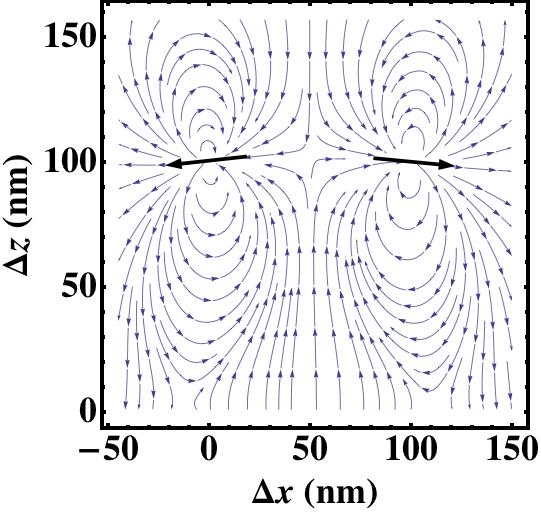}
\caption{\label{fig:dipolerad} 
Electric field lines in the near field of two dipoles driven on the 
primary resonance (a), and the secondary resonance (b) near a metal interface at $z =0$. 
The bold arrows indicate the dipole location and orientation of the emitters.}
\end{center}
\end{figure}

Figure~\ref{fig:mirintresp}(b) indicates that at the secondary resonance the dipole orientation
changes from normal to almost parallel to the interface.
Because the two dipoles are nearly anti-parallel to each other their emitted radiation
interferes destructively. This implies that the system changes from
superradiant behaviour around the primary resonance to a subradiant behaviour
around the secondary resonance. Figure~\ref{fig:dipolerad} displays the electric near-field
and the dipole orientation at the two resonances. An intuitive explanation of the secondary resonance can be given as follows. For most
driving field frequencies the coupling between the $x$ and $z$ components of the oscillators
is weak. However, at the resonance frequency of sub-radiant modes this coupling can 
transfer most of the energy of the oscillator from the $z$ to the $x$ components because
the latter only lose very little energy energy through radiative emission and thus decay
slowly. The secondary resonance therefore has to be narrow because its linewidth
is inversely proportional to the large decay time of the $x$-oscillations. 

To improve our understanding we compare in Figs.~\ref{fig:mirintresp}(a) and (b), a numerical evaluation of the oscillation amplitudes for two emitters near the metal
film with a corresponding calculation for two emitters near a perfect mirror. 
The latter case can be described  by taking the dielectric constant $\epsilon_2$ of the metal film
to be infinite, which corresponds to a perfect conductor.
While the mirror images explain the qualitative features very well, 
the narrow resonance is greater in the case of the mirror. We attribute this difference to the existence of surface plasmons, which are 
not present in the case of a perfect mirror. The dispersion and decay of surface plasmons cause the emitter energy to be dissipated quicker and this is represented by a corresponding decrease and broadening in the narrow resonance.

With increasing distance from the interface the narrow resonance falls off exponentially 
because of the decreasing strength of the reflected near field at the interface.
Presumably for the same reason the resonance frequencies are shifted by a larger amount near the metal film.

\section{Collective decay of classical emitters} \label{sec:decay}

We model the collective decay of initially excited oscillators
near the interface
by assuming that there is no driving field and that
the emitters perform a simple damped harmonic motion
$\bm{x}^{(j)}(t)=\bm{x}^{(j)}(0)e^{-i\omega t-\Gamma t}$,
with $\omega$ and $\Gamma$ the collective oscillation frequency
and decay rate, respectively. Using the results of 
appendix \ref{app:fieldEqn} again we can transform
Eq.~(\ref{eqn:coupDE}) into a transcendental equation
for the collective parameters $\omega$ and $\Gamma$,
\begin{align}
\label{algEq1a}
  \left[(\Gamma + i\omega)^2 + \omega_0^2
  \right]\bm{x}^{(j)}(t)
  &= 
  - \frac{\mu_0 q^2}{m}  (\Gamma+i\omega)^2 
   \sum_{i=1}^N  \bm{x}^{(i)}(t) 
\nonumber \\  &\times 
 \bm{G}(\bm{r}^{(j)}, \bm{r}^{(i)},
  \omega-i\Gamma).
\end{align}

For optical emission it is safe to assume that 
$\Gamma \ll \omega$,
i.e.,  the duration of the emitted light pulse is
much longer than the inverse frequency. Because 
$\bm{G}(\bm{r}^{(j)}, \bm{r}^{(i)}, \omega)$, is a meromorphic
function of $\Omega$ in the lower half plane, and because 
its poles are determined by the properties of the metal
interface and are unrelated to $\Gamma$, we can approximate
$\bm{G}(\bm{r}, \bm{r}^{(j)},\omega-i\Gamma)$
by $\bm{G}(\bm{r}, \bm{r}^{(j)}, \omega -i\epsilon)$,
where $\epsilon >0$ is infinitesimally small.
This reduces the transcendental equation (\ref{algEq1a})
to a linear eigenvalue problem 
\begin{equation}
\label{algEq2}
\left [(i\Gamma +\delta)\openone - \frac{\mu_0 q^2 \omega_0}{2
    m}\sum_{i=1}^N\bm{G}(\bm{r}^{(j)}, \bm{r}^{(i)},
  \omega) \right]\bm{x}^{(i)}(0)= 0,
\end{equation}
with frequency shift $\delta \equiv \omega_0 -\omega$ $\ll \omega_0$.
The decay parameter $\Gamma$ and the  frequency shift $\delta$
relate to the imaginary and real part of the eigenvalue problem (\ref{algEq2}), respectively.
Generally, $\Gamma$  characterizes the collective decay rate of the
all oscillators.

\subsection{Decay rate}
 It is instructive to first study the
decay a single oscillator $\bm{x}(t)$ at position $\bm{r}$.
In free space the Green's function takes the form \cite{KSW} 
\begin{equation}
 \bm{G}(\bm{r}, \bm{r}, \omega)
  = \left ( G' + i \frac{\omega_0}{6 \pi
    c} \right ) \openone.
\end{equation}
The real part $G'$ is related to the Lamb shift of atomic
resonance lines. It is formally divergent and would need
to be renormalized \cite{brooke08}. However, because we are only
interested in the decay rate, we can ignore the line shift
and assume it is absorbed in the definition of the detuning.
Because the Lamb shift is much smaller than the optical resonance
frequency this will result in an excellent approximation
for the decay rate. As in free space the matrix in Eq.~(\ref{algEq2})
is proportional to the identity the corresponding eigenvalue
problem is trivially solved and yields the 
well-known decay rate of a single oscillator,
\begin{align}
\label{eqn:totaldecay}
\Gamma_0  &= \frac{\mu_0 q^2 \omega_0}{2 m} 
 \mathrm{Im}G_{\mu\mu}(\bm{r}, \bm{r}, \omega_0)
\\  &=  
\frac{\mu_0 q^2 \omega_0^2}{12 \pi m c}.
\end{align}

\begin{figure}
\begin{center}
\includegraphics[width=6cm]{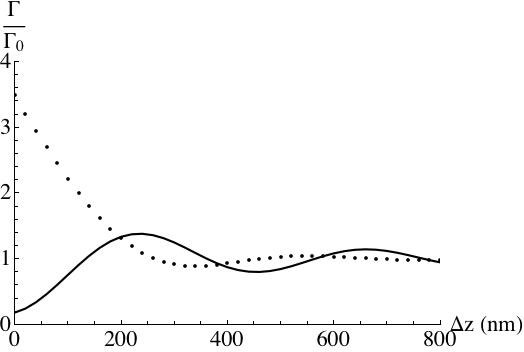}
\caption{\label{fig:decaypic} The decay rate of a single oscillator
  placed within the vicinity of a thin metal film, at a separation
  distance of $\Delta z$. the decay rates are given for a dipole
  oriented parallel (solid) and perpendicular (dotted) relative to the
  plane of the interface.}
\end{center}
\end{figure}

The decay rate near the interface can be derived in a similar way
as long as $\Gamma$ is much smaller than the optical frequency and
varies little for frequency variations on the order of the Lamb shift.
These assumptions should be satisfied as long as the emitter is not
too close to the interface. We remark, however, that these approximations 
are not universally valid
and fail in the case of photonic band gap materials
\cite{john94}, for instance. 
The Green's function 
$G_{\mu\nu}(\bm{r}, \bm{r}, \omega)$ evaluated
at a single point $ \bm{r}$ near the interface is diagonal 
in Cartesian coordinates
(cf. App.~\ref{sec:greenf}).
We therefore can use Eq.~(\ref{eqn:totaldecay}) to find the decay rate.

Fig.~\ref{fig:decaypic}  depicts the decay rate of an emitter as a function of
the distance $\Delta z$ of the emitter from the metal film
(cf.~Fig.~\ref{fig:interface}). It demonstrates that the decay
rate is enhanced (suppressed) if the emitter oscillates perpendicular
(parallel) to the interface, respectively. 
This can be understood using the concept of mirror
images with each emitter considered as an electric dipole (see App.~\ref{app:fieldEqn}).
If a dipole is oriented perpendicular to the interface, its mirror image is in phase and thus can
enhance the emission of the dipole. 
On the other hand, dipoles that oscillate in the plane of the interface have mirror images 
that are 180$^\circ$ out of phase so that the radiation reaction causes damping of the dipole. 
Hence, if the mirror images were composed of real charges these two situations would just correspond to superradiant
and subradiant collective emission of two emitters, respectively.

\begin{figure}
\includegraphics[width=4cm]{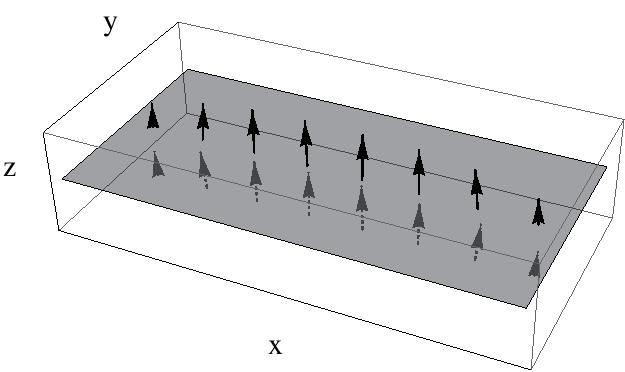}
\includegraphics[width=4cm]{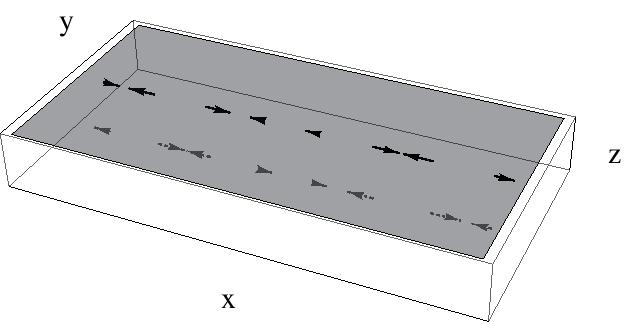}
\centerline{(a)\hspace{3.7cm}(b)}
\\
\includegraphics[width=4cm]{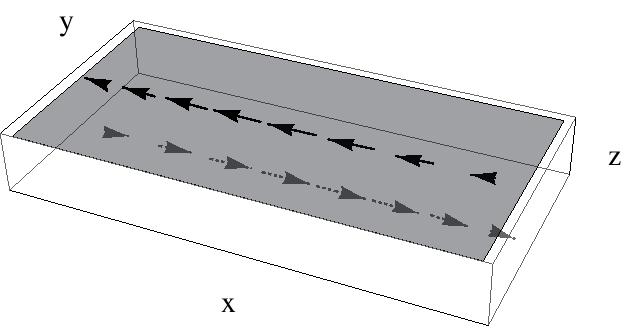}
\includegraphics[width=4cm]{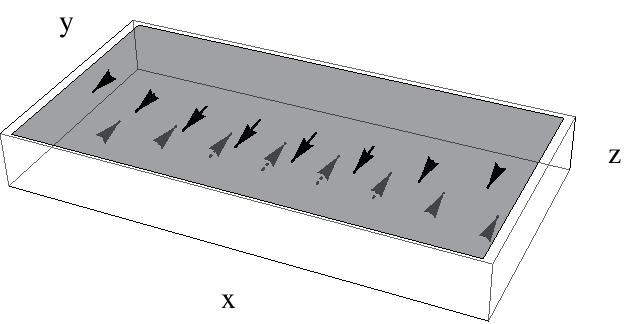} 
\centerline{(c)\hspace{3.7cm}(d)}
\caption{\label{fig:oscillPics} 
Four different collective eigenmodes: (a) superradiant mode, (b) subradiant mode, (c)  and (d) 
are suppressed superradiant
modes for which all real dipole moments are in phase but all mirror images are out of phase.
Shown is the orientation of the dipole moment for real emitters (above the interface) and their mirror images (dashed, below
the interface).} 
\end{figure}

We now turn to the
collective decay of $N>1$ emitters. 
In this case there are $3N$ eigenstate solutions of Eq.~(\ref{algEq2}).
Fig.~\ref{fig:oscillPics} displays four out of 24 numerically determined eigenmodes for $N=8$ which each are examples
for a particular collective behaviour.
In free space a state where all emitters oscillate in phase and in the same direction is superradiant. 
The presence of the interface breaks the axial symmetry and oscillations in the plane of 
the interface are suppressed because the corresponding mirror images trigger subradiant behaviour.
Fig.~\ref{fig:oscillPics}(a) shows the superradiant mode where the emitters oscillate
in phase along the z-axis. In this case their mirror images are also in phase and enhance the
radiation intensity. Fig.~\ref{fig:oscillPics}(b) shows a subradiant mode where the emitted
fields of the individual emitters interfere destructively. Fig.~\ref{fig:oscillPics}(c) and (d) display
a very different state which we call suppressed superradiant states. All oscillators are in phase
and in free space it would be a superradiant state. However, the mirror images are out
of phase so that the intensity of the emitted radiation is strongly reduced as compared to
the superradiant state. In Fig.~\ref{fig:radScaling}(a), which will be discussed below,
we display the collective decay rate of these modes
as a function of the number of oscillators.

\subsection{Far-field radiation}
A simple way to learn about the dynamics of a collection of emitters
is to observe their emission pattern in the far field. 
The radiation intensity is determined by the (time averaged) Poynting vector 
\begin{equation}
\label{eqn:Pvec}
\bm{S}(\bm{r,t})=\frac{1}{2\mu_0c}|\bm{E}(\bm{r,t})|^2\hat{\bm{r}}.
\end{equation}
The electric field of each emitter is determined by Eq.~(\ref{EfieldJ}) and the total
field $\bm{E}(\bm{r,t}) $ is the superposition of the individual fields. To evaluate Eq.~(\ref{EfieldJ})
in the far field we have to compute the radiative Green's function (\ref{eqn:FTGreen})
in the far-field limit, $kr \rightarrow \infty$,
which can be accomplished using the method of stationary phase
described in Appendix \ref{sec:statphase}.

For a source in the region
$z'>d$ and observation of the field at position $\bm{r}$ in the region
$z<0$,
 the dyadic Green's function can be approximated as 
\begin{align} 
\label{eqn:statphaseGxx}
  G_{\mu\nu}(\bm{r},\bm{r}',\omega) &= 
  \frac{i}{2\pi}\left(\frac{k_1}{r}\frac{z}{r}\right)e^{ik_1r}e^{-ik_1\sin{\phi}(\cos{\theta}x'+\sin{\theta}y')}
\nonumber  \\ & \times
  \tilde{G}_{\mu\nu}(k_1\sin{\phi}\cos{\theta},k_1\sin{\phi}\sin{\theta},\omega;z'),
\end{align} 
with the unwieldy coefficients $\tilde{G}_{\mu\nu}$ defined in Eqs.~(\ref{G13A}) - (\ref{G13B}). A similar expression can be derived for an observation point in region 3.
The direction from the emitters to the observation point is given by
$(\sin{\phi}\cos{\theta}, \sin{\phi}\sin{\theta}, \cos\phi)$.
The radiation profiles are determined from the positive frequency
components of the electric field in the far field. For emitters that
are equidistant from the interface, the Poynting vector can be
represented as  
\begin{equation} 
\label{eqn:ffPvec}
\begin{split}
  |\bm{S}(\bm{r,t})| =& 
  \frac{\mu_0 \omega^4}{8\pi^2 c}\left( \frac{k z}{r^2} \right)^2\\
  & \times
  |\tilde{G}_{\mu\nu}(k_i\sin{\phi}\cos{\theta},k_i\sin{\phi}\sin{\theta},\omega;z')|^2\\
  & \times
  \left|\sum_i^N
  e^{-ik\sin{\phi}(\cos{\theta}x^{(i)}+\sin{\theta}y^{(i)})}e^{-\Gamma
    t}x^{(i)}\right|^2,
\end{split}
\end{equation} 
with $j=1, 3$ for the observation point in region 1 or 3, respectively.

The last term is the relevant phasing term which determines
the $N^2$-gain in intensity that is associated with superradiance. 
The superradiant decay modes must be in phase, i.e.,
$k\sin{\phi}(\cos{\theta}x^{(i)}+\sin{\theta}y^{(i)}) + \text{arg}(x^{(i)}_\mu) =n_i 2\pi$ for each emitter $i$,
then the sum becomes proportional to $N^2$. In other words, constructive interference between the emitted radiation of all emitters 
is one condition for observing superradiance.

\begin{figure}
\begin{center}
\includegraphics[width=6cm]{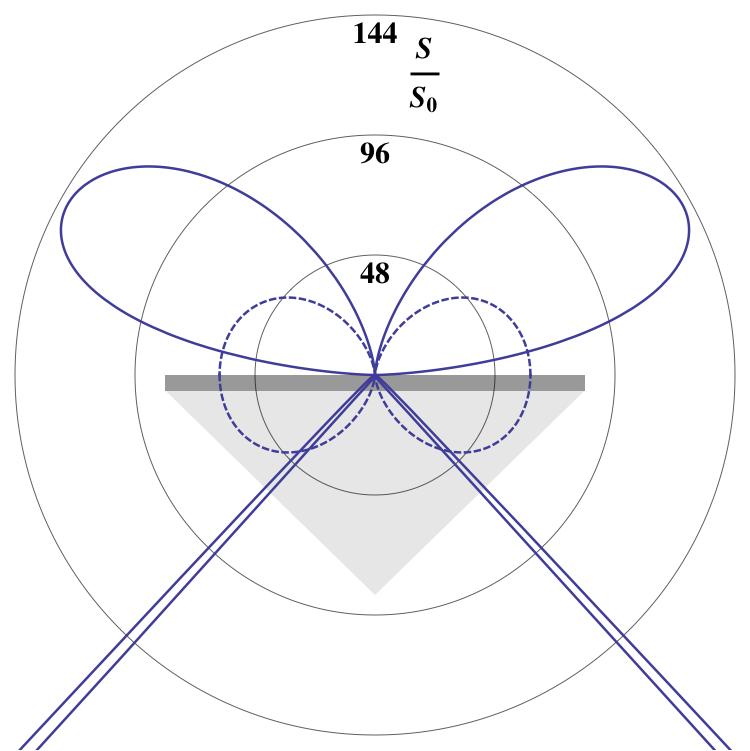}
\caption{\label{fig:transamp} 
The far-field emission pattern of 8 oscillators in the superradiant mode
at a distance of $\Delta z = 200$nm from the interface (solid)
and in free space (dashed). Shown is the ratio of the radiation intensity $S$
 and the maximum intensity $S_0$ produced by a single emitter in free
  space. The emitters are placed parallel to the horizontal axis.
} 
\end{center} 
\end{figure}

The transmitted radiation profile (\ref{eqn:ffPvec}) 
for the superradiant state in the plane of the emitters ($\theta=0$)
is illustrated in Fig.~\ref{fig:transamp} for a linear arrangement of 
8 emitters periodically separated within a distance of $\lambda/2$,
with $\lambda=800$nm and $\omega = 2\pi c/\lambda$.
The dashed curve shows the emission pattern in free space.
For perfect superradiance its maximum intensity (along the horizontal axis)
should scale like $S= A N^2 S_0$, with
$S_0$ is the maximum intensity for a single emitter in free space
and with the superradiance prefactor $A=1$  in free space.
Because the outermost oscillators are a distance $\lambda/2$ apart so that
their emitted fields are out of phase, the $N^2$ scaling is only roughly fulfilled.

For emitters near the interface (solid curve in Fig.~\ref{fig:transamp})
surface plasmons lead to the narrow but extremely large peaks in the lower half of  Fig.~\ref{fig:transamp}
at the plasmon resonance angle with a width of about $0.01$ radian.
Emission into the radiative modes around this peak is superradiant (see below) with $A\approx 240$. 
The value of $A$ depends on the separation distance of the emitters from the interface and the 
Fresnel coefficient $\tilde{T}_{3,2}^{\text{TM}}$ defined in Appendix~\ref{sec:greenf}.
 $A$ will decrease exponentially as the separation distance between emitter and interface increases due to decreased near field coupling of the emitter to surface plasmon modes. For given dielectric constants $\varepsilon_i$ and emission wavelength $\lambda$ 
there is an optimum film thickness for which $A$ is maximized.
The values used in this paper correspond to an optimum film thickness of $56$nm for $\lambda=800$nm.
Hence, the radiation from multiple emitters transmitted
into the surface plasmon resonance angle  (the spikes in Fig.~\ref{fig:transamp})
combines the enhancement $A$  and the $N^2$ gain of
superradiance. 
For the case $N=8$ the peaks have a maximum of almost 15000 times
the maximum intensity $S_0$  for a single emitter in free space.

\subsection{Suppressed superradiance}

\begin{figure}
\begin{center}
\centerline{{\bf \hspace{0.4cm}(a)\hspace{3.5cm} (b)}}
\includegraphics[width=3.9cm]{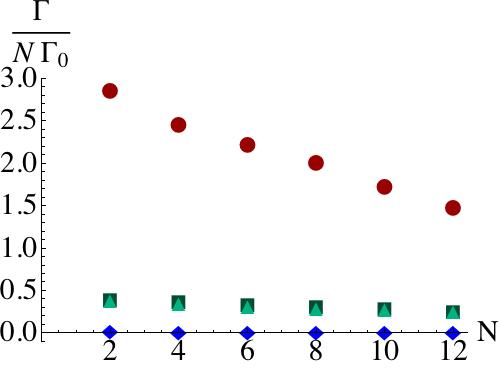}
\includegraphics[width=3.9cm]{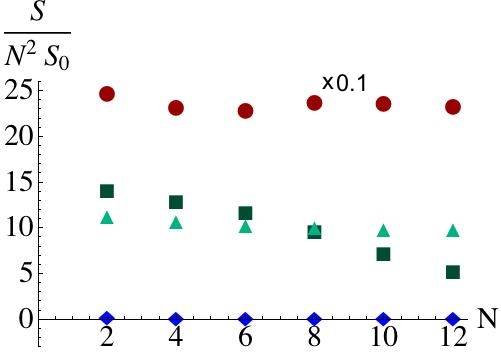}
\caption{\label{fig:radScaling} (a) Scaling of the collective decay rate with the number $N$ of oscillators.
Circles correspond to the superradiant mode, squares and triangles to the suppressed superradiant modes
displayed in Fig.~\ref{fig:oscillPics} (c) and (d), respectively, and
diamonds to the subradiant mode displayed in Fig.~\ref{fig:oscillPics} (b).
The distance between adjacent oscillators is 50 nm. Each oscillator is a distance $\Delta z=100$nm away
from the interface.
In (b) the scaling of the maximum intensity of the peak at the surface
plasmon resonance angle with the number $N$ of oscillators. In the superradiant case the results are reduced by 
a factor of 10 for presentational purposes.}
\end{center}
\end{figure}

We conclude this section with a discussion of the main indicators for superradiance
in the superradiant and the suppressed superradiant states.
In Fig.~\ref{fig:radScaling}(a) we display the scaling of the collective decay parameter with the number of oscillators, which is a measure for how strongly light emitted from one oscillator can drive emission from another oscillator. Because of the growing size of the arrangement of oscillators, which varies from 50 nm for $N=2$ to 550 nm for $N=12$, we expect superradiant phenomena to decrease with $N$. 
This is indeed the case for the superradiant state (circles), but for the
suppressed superradiant states collective decay ($\Gamma \sim N$) (triangles and squares) is preserved.

Interestingly, we observe the opposite situation with respect to the $N^2$ dependence
of the peak intensity at the surface plasmon resonance angle, 
which is shown in Fig.~\ref{fig:radScaling}(b). 
For the superradiant state
the peak intensity can well be described by $S \approx A N^2 S_0$,
indicating constructive interference of the emitted radiation despite the growing size
of the sample. For the suppressed superradiant states the ability to constructively 
interfere depends on the dipole orientation: the $N^2$ scaling is significantly affected
only if the dipoles are aligned as in Fig.~\ref{fig:oscillPics}(c).

Clearly the two suppressed superradiant states fulfill the criteria for superradiance
as well, or better, than the superradiant state itself. We use the notion ``suppressed''
to characterize their behaviour because their decay rate is reduced by a factor of 6 and their
peak intensity by a factor of 25 as compared to the superradiant state.
We can understand suppressed superradiant behavior by considering
each dipole and its mirror image as one entity. The dipole moment of this entity vanishes, but it
can emit (a weaker) quadrupole radiation. Because all $N$ quadrupoles are in phase their
emitted fields interfere constructively and the intensity scales like $N^2$. Furthermore, similar to how a neighbouring dipole in phase can increase the decay rate, a quadrupole can drive the neighbour when in phase. Hence, the suppressed superradiant
state can be considered a superradiant state for quadrupoles. 

The different behaviour of the states of Fig.~\ref{fig:oscillPics} can be understood by considering the
mirror images as real charges. For suppressed superradiant states the spatial size of each quadrupole, 
$2\Delta z = 200$nm, is larger
than the distance $\Delta x=50$nm between adjacent quadrupoles.
Hence, they interact with their respective near field so that cooperative decay can still grow like $N$. 
On the other hand,
for the superradiant state each dipole interacts with the far field of other dipoles so that the
deviation from the linear scaling with $N$ is more pronounced, albeit the overall interaction energy
is much larger.

The $N^2$ scaling of the peak intensity can be understood by taking into account
that the maximum peak appears at the surface plasmon resonance angle, i.e., 
in a direction that is different from the axis along which the oscillators are aligned.
Generally, surface plasmons can only be generated in the direction of the polarization of
the emitted radiation because it is the latter's electric field that generates electron density oscillations.
For the suppressed superradiant  state \ref{fig:oscillPics}(c) there are significant deviations
from the $N^2$ law because the dipoles oscillate in the plane of the interface parallel to the $x$-axis.
This means that most radiation is emitted into a surface plasmon mode that propagates
in the $x$-direction, but in this direction phase variations have a very pronounced
effect. 
For the suppressed superradiant  state \ref{fig:oscillPics}(d), as for the superradiant state,
most radiation is emitted into a plasmon that propagates in the $y$-direction, i.e.,
parallel to the alignment of emitters.
Therefore their emitted radiation fields can still constructively interfere.

\section{Conclusion}

In this paper we give a detailed account of how super- and sub-radiant
emission by oscillating dipoles is affected by surface plasmons. It was shown
that the interface generates an indirect coupling between the emitters through
light that is reflected by the interface.
 In the superradiant mode the peak intensity of the emitted radiation 
is two orders of magnitude larger than a similar arrangement of emitters in free-space.
The decay rate of the system is increased by the presence of mirror dipoles at the interface. 
Applying a driving field to the emitters can induce super- and sub-radiant emitter modes by a suitable choice of frequency.

The collective decay eigenmodes of initially excited oscillators showed that certain
modes, which naively would be expected to behave in a sub-radiant fashion, are
actually superradiant modes, albeit with a strongly suppressed overall intensity.
These suppressed superradiant modes are also a consequence of the additional
coupling between oscillators that is generated by the interface. An intuitive picture
explains this effect as superradiance of quadrupole radiation.

Our results will help to improve the characterization of the radiative response of
entangled atoms or ions in quantum information and of surface-enhanced
spectroscopy of molecular excitations. A detailed study of photon entanglement
in the presence of surface plasmons is currently under investigation. 

\acknowledgments

The authors are indebted to Ren\'e Stock for many
enlightening discussions about surface plasmons
from which this work has greatly benefitted.
Funding by NSERC, iCORE, and QuantumWorks
is gratefully acknowledged. J.~J.~C.~is grateful for the support from NSERC RCD. B.~C.~S.~is a CIFAR Associate.

\appendix

\section{Derivation of the dynamical equation}\label{app:fieldEqn}
The propagation of an electric field $\bm{E}(\bm{r}, t)$ in
the presence of multiple interfaces is solved using a Green's Function
approach. The electric field due to a current source of
$\bm{j}(\bm{r}, t)$ is determined by the well known
differential wave equation 
\begin{equation}
\label{eqn:EfieldDE}
\left(\frac{1}{c^2}\varepsilon(\bm{r})\left(\pd{}{t}\right)^2
  + \nabla\times\nabla\times \right)\bm{E}(\bm{r}, t) =
-\mu_0\pd{}{t}\bm{j}(\bm{r}, t), 
\end{equation}
where $\varepsilon(\bm{r})$ is the complex relative
permittivity of a dielectric. 
We assume here for simplicity that the permittivity of the interfaces
does not depend on the radiation frequency, which is the case for the
interface that we are studying
\footnote{
We remark that our methods have a much broader range of
validity. The central equations (\ref{eqn:n3osc}) and (\ref{algEq2})
of our investigation can also be derived
using a temporal Laplace transform by exploiting the relation
$\text{FT}(f)(\omega) = \lim_{\eta\rightarrow 0} \text{LT}(f)(\eta
-i\omega)$ between the Fourier transform FT of a function $f(t)$
and its Laplace transform LT. 
Eq.~(\ref{eqn:n3osc}) is then valid for an arbitrary frequency dependence
of the permittivity.
Eq.~(\ref{algEq2}) holds for a frequency dependend permittivity
provided the Green's function varies little over the frequency range 
of the radiation pulse. This is the case if the medium is not resonant
within this frequency range.
}. The constants $c$ and $\mu_0$ are the
speed of light and permeability of free space respectively. The
solution to Eq.~(\ref{eqn:EfieldDE}) in terms of the dyadic Green's
function $\bm{G}(\bm{r}, t; \bm{r}', t')$ is 
\begin{equation}
\label{eqn:EfieldGF}
  \bm{E}(\bm{r}, t) =
  -\mu_0{\iint}dt'd^3\bm{r}'
  \bm{G}(\bm{r}, t; \bm{r}',t')
  \pd{}{t'}\bm{j}(\bm{r}', t'), 
\end{equation}

We assume that each emitter consists of a charge $q$ (the ``electron'') 
that performs small oscillations of amplitude $\bm{x}^{(j)}(t)$ around a fixed position $\bm{r}^{(j)}$
at which a charge $-q$ (the ``nucleus'') is placed.
The current density $\bm{j}(\bm{r}',t')$ produced by
emitter $j$ can then be written as 
\begin{equation}
\label{eqn:oscden}
\bm{j}(\bm{r}',t')=q\delta(\bm{r}'-\bm{r}^{(j)})\dot{\bm{x}}^{(j)}(t').
\end{equation}
This current density contains the complete information about the radiation field produced by both
the electron and the resting nucleus and determines the radiative electric field through Eq.~(\ref{eqn:EfieldGF}).
In addition one generally needs to include the electric Coulomb field of a charge distribution. However,
because the oscillation amplitudes of the electron are small and the total charge of the emitter is zero,
each oscillator essentially describes a point dipole with dipole moment $\bm{p} = q \bm{x}(t)$.
We then can ignore the Coulomb contribution so that each emitter can be thought of as an oscillating
point dipole.
Inserting Eq.~(\ref{eqn:oscden}) into Eq.~(\ref{eqn:EfieldGF}) then yields
the electric field (\ref{EfieldJ}) produced by emitter $j$.

The radiative Green's function 
$\bm{G}(\bm{r}, t; \bm{r}', t')$ is a retarded solution to
\begin{eqnarray} 
\label{eqn:GFDE}
\left(\frac{1}{c^2}\varepsilon(\bm{r})\left(\pd{}{t}\right)^2
  + \nabla\times\nabla\times \right)\bm{G}(\bm{r}, t;
\bm{r}', t') 
 & &
\nonumber \\ 
=
\bm{I}\delta(\bm{r}-\bm{r}')\delta(t-t').  & &
\end{eqnarray} 
The Green's functions in frequency domain and Cartesian coordinates are
derived from the equation 
\begin{eqnarray} 
\label{eqn:dyadicDE}
\left[\frac{\partial}{\partial{r_\gamma}}\frac{\partial}{\partial{r_\mu}}-\delta_{\gamma\mu}\left(\Delta+\frac{\omega^2}{c^2}\epsilon(\bm{r},
    \omega)\right)\right]\GF{\mu}{\nu}
 & &
\nonumber \\ 
=\delta_{\gamma\nu}\delta(\bm{r}-\bm{r}'). & &
\end{eqnarray} 
where the indices indicate the appropriate cartesian
coordinates. Considering that the sources are located in region 3, the
solution to Eq.~(\ref{eqn:dyadicDE}) is \cite{DKW}, 
\begin{equation}
\label{eqn:gfsource}
\GF{\mu}{\nu}=\left[k_3^{-2}\frac{\partial}{\partial{r_\mu}}\frac{\partial}{\partial{r_\nu}} 
  + \delta_{\mu\nu}
\right]\frac{e^{ik_3|\bm{r}-\bm{r}'|}}{4\pi
  |\bm{r}-\bm{r}'|}, 
\end{equation}
where $k_3=\sqrt{\varepsilon_3}\frac{\omega}{c}$. The Green's functions
corresponding to the reflected and transmitted fields are obtained by
solving the homogeneous equation 
\begin{equation}
\left[\frac{\partial}{\partial{r_\gamma}}\frac{\partial}{\partial{r_\mu}}-\delta_{\gamma\mu}\left(\Delta+\varepsilon_3\frac{\omega^2}{c^2}\right)\right] 
{G_{\mu\nu}(\bm{r}, \bm{r}', \omega)} = 0.  
\end{equation}
The solution to these equations with applied boundary conditions are
found in Appendix \ref{sec:greenf}. 

%%%%%%%%%%%%%%%%%%%%%%%%%%%%%%

Let us now assume that emitter $j$ oscillates according to
\begin{equation}
\label{eqn:decayamp}
\bm{x}^{(j)}(t)=\bm{x}^{(j)}(0)e^{-i\omega t-\Gamma t}
\end{equation}
for some frequency $\omega$
and a real, positive decay parameter $\Gamma \ll \omega$. 
Eq.~(\ref{EfieldJ}) then becomes
\begin{eqnarray}
  \bm{E}^{(j)}(\bm{r}, t) &=&
  -\mu_0 q \int_{-\infty}^{\infty}dt' 
   \bm{G}(\bm{r}, \bm{r}^{(j)}, t-t')
\nonumber \\ & & \times
   (\Gamma+i\omega)^2\bm{x}^{(j)}(0)e^{-i\omega t' -\Gamma t'}
\nonumber\\ & = &  
  -\mu_0 q 
  (\Gamma+i\omega)^2\,  e^{-i\omega t -\Gamma t}
\nonumber \\ & & \times
  \bm{G}(\bm{r}, \bm{r}^{(j)}, \omega -i\Gamma)
\, \bm{x}^{(j)}(0)  \; ,
\end{eqnarray}
where $ \bm{G}(\bm{r}, \bm{r}^{(j)}, \omega -i\Gamma)$ is
the analytic continuation of the temporal Fourier transform of 
 $ \bm{G}(\bm{r}, \bm{r}^{(j)}, t)$ into the lower half plane.

Noting the pole at $\omega_1=\omega-i\Gamma$ and applying calculus of residues, the electric field is then
\begin{equation}
\label{eqn:EfieldSrc}
\bm{E}^{(j)}(\bm{r}, t) = -\mu_0 q\bm{G}(\bm{r},
\bm{r}^{(j)}, \omega-i\Gamma)
(\Gamma+i\omega)^2\bm{x}^{(j)}(0)e^{-i\omega t-\Gamma t}. 
\end{equation}
Inserting Eq.~(\ref{eqn:EfieldSrc}) and Eq.~(\ref{eqn:decayamp}) into Eq.~(\ref{eqn:coupDE}) gives
\begin{eqnarray} 
  \left[(\Gamma + i\omega)^2 + \omega_0^2
  \right]\bm{x}^{(j)}(t)
  &=&
   \frac{q}{m}  \Big \{ \bm{E}(\bm{r}^{(j)},t) -\mu_0 q
\nonumber \\ &  \times &
  \sum_{i=1}^N\bm{G}(\bm{r}^{(j)}, \bm{r}^{(i)},
  \omega-i\Gamma)
\nonumber \\ &  \times &
   (\Gamma+i\omega)^2  \bm{x}^{(i)}(t) \Big \}
\label{algEq1}\end{eqnarray} 

\section{Green's Functions of a 3-layer dielectric}
\label{sec:greenf}

Using the method of Ref.~\cite{maradudinmills} (See also Ref.~\cite{DKW_SI}) the
dyadic Green's functions of a multiple layered planar dielectric are
solved. Due to the translational invariance of the problem of a planar
dielectric interface, it is useful to decompose the Green's function
into transverse and normal components through the Fourier transform 
\begin{equation}
\label{eqn:FTGreen}
  \GF{\mu}{\nu} = \int{\frac{d^2\bm{k}_p}{(2\pi)^2}
  e^{i\bm{k}_p\cdot(\bm{r}_p-\bm{r}_p')}\fGF{\mu}{\nu}}, 
\end{equation}
where $\bm{k}_p$ is the wave vector component tangential to the
interface and $\bm{r}_p$ is the corresponding position component. 

For sources located in region 3 ($z'>d$) and $z>d$ the elements of
$\fGF{\mu}{\nu}$ are 
\begin{eqnarray} 
  \fGFZ{x}{x} &=& \frac{1}{\bm{k}_p^2} 
  e^{-i \beta_3 (z+d)}
  \Big ( -k_x^2 \tilde{R}^{\text{TM}}_{3,2} Z_{xx}(\bm{k}_p,z') 
\nonumber \\ & & \hspace{1cm}
   + k_y^2 \tilde{R}^{\text{TE}}_{3,2} Z_{yy}(\bm{k}_p,z') 
   \Big ) 
\\ 
  \fGFZ{x}{y} &=& -\frac{k_x k_y}{\bm{k}_p^2}
  e^{i \beta_3 (z-d)}
  \Big(
    \tilde{R}^{\text{TM}}_{3,2} Z_{xx}(\bm{k}_p,z') 
\nonumber \\ & & \hspace{1cm}
   + \tilde{R}^{\text{TE}}_{3,2} Z_{yy}(\bm{k}_p,z') 
  \Big)
\\
  \fGFZ{x}{z} & = & 
  -\frac{k_x}{\bm{k}_p} \tilde{R}^{\text{TM}}_{3,2} 
  Z_{xz}(\bm{k}_p,z')e^{i \beta_3 (z-d)}
\\
  \fGFZ{y}{x} & = & \fGFZ{x}{y}
\\
  \fGFZ{y}{y} &=&  \fGFZ{x}{x}, \phantom{hh}k_x\leftrightarrow k_y
\\
  \fGFZ{y}{z} &=& \fGFZ{x}{z}, \phantom{hh}k_x\leftrightarrow k_y
\\
  \fGFZ{z}{x} &=& 
   \frac{k_x}{\beta_3} \tilde{R}^{\text{TM}}_{3,2} Z_{xx}(\bm{k}_p,z')
  e^{i \beta_3 (z-d)}
\\
  \fGFZ{z}{y} &=& \frac{k_y}{\beta_3} 
  \tilde{R}^{\text{TM}}_{3,2} Z_{xx}(\bm{k}_p,z')
  e^{i \beta_3 (z-d)}
\\
  \fGFZ{z}{z} &=& \frac{k_p}{\beta_3} 
  \tilde{R}^{\text{TM}}_{3,2} Z_{xz}(\bm{k}_p,z')
  e^{i \beta_3 (z-d)}
\end{eqnarray} 
and for $z<0$ the terms are
\begin{eqnarray} 
  \fGFZ{x}{x} &=& \frac{1}{\bm{k}_p^2}
  e^{-i \beta_1 z}
  \Big ( k_x^2 \tilde{S}^{\text{TM}}_{3,1} Z_{xx}(\bm{k}_p,z')
\nonumber \\ & & \hspace{1cm}
  + k_y^2
  \tilde{S}^{\text{TE}}_{3,1} Z_{yy}(\bm{k}_p,z') 
  \Big)
\label{G13A}\\
  \fGFZ{x}{y} &=& \frac{k_x k_y}{\bm{k}_p^2}
  e^{-i \beta_1 z }
  \Big( \tilde{S}^{\text{TM}}_{3,1} Z_{xx}(\bm{k}_p,z')
\nonumber \\ & & \hspace{1cm}
  - \tilde{S}^{\text{TE}}_{3,1} Z_{yy}(\bm{k}_p,z') \Big)
\\
  \fGFZ{x}{z} &=& \frac{k_x}{\bm{k}_p} 
  \tilde{S}^{\text{TM}}_{3,1} Z_{xz}(\bm{k}_p,z'e^{-i \beta_1 z})
\\
  \fGFZ{y}{x} &=& \fGFZ{x}{y}
\\
  \fGFZ{y}{y} &=& \fGFZ{x}{x}, \phantom{hh}k_x\leftrightarrow k_y
\\
  \fGFZ{y}{z} &=& \fGFZ{x}{z}, \phantom{hh}k_x\leftrightarrow k_y
\\
  \fGFZ{z}{x} &=& \frac{k_x}{\beta_3} \tilde{S}^{\text{TM}}_{3,1} 
  Z_{xx}(\bm{k}_p,z')e^{-i \beta_1 z }
\\
  \fGFZ{z}{y} &=& \frac{k_y}{\beta_3} \tilde{S}^{\text{TM}}_{3,1} 
  Z_{xx}(\bm{k}_p,z')e^{-i \beta_1 z}
\\
  \fGFZ{z}{z} &=& \frac{k_p}{\beta_3} \tilde{S}^{\text{TM}}_{3,1} 
  Z_{xz}(\bm{k}_p,z')e^{-i \beta_1 z}
\label{G13B}\end{eqnarray} 
where
\begin{eqnarray} 
  Z_{xx}(\bm{k}_p,z') &=&
  \frac{i\beta_3}{2 k_3^2}e^{i \beta_3 |d-z'|}
\\
  Z_{yy}(\bm{k}_p,z') &=&
  \frac{i}{2 \beta_3}e^{i \beta_3 |d-z'|}
\\
  Z_{xz}(\bm{k}_p,z') &=& \frac{i k_p}{2 k_3^2}e^{i \beta_3 |d-z'|}\; ,
\end{eqnarray} 
with $\beta_i=\sqrt{k_i^2-k_p^2}$ and $k_i=\sqrt{\varepsilon_i}\frac{\omega}{c}$.

The transmission and reflection of radiation at the interface is
characterized by the generalized Fresnel reflection and transmission
coefficients. Consider a series of interfaces, where $i$ increases for
$z$ increasing in $z>0$. For reflection in region $i$, from an
interface between regions $i$ and $i-1$, the generalized reflection
coefficient is 
\begin{equation}
\tilde{R}^{\text{TM/TE}}_{i,i-1}=\frac{R^{\text{TM/TE}}_{i,i-1}+\tilde{R}^{\text{TM/TE}}_{i-1,i-2}e^{i2\beta_{i-1}(d_{i-1}-d_{i-2})}}{1+R^{\text{TM/TE}}_{i,i-1}\tilde{R}^{\text{TM/TE}}_{i-1,i-2}e^{i2\beta_{i-1}(d_{i-1}-d_{i-2})}}.
\end{equation}
The generalized transmission coefficients are
\begin{equation}
\tilde{T}^{\text{TM/TE}}_{i,j}=\prod_{m=j+1}^{i} \frac{T^{\text{TM/TE}}_{m,m-1}e^{i\beta_m(d_m-d_{m-1})}}{1+R^{\text{TM/TE}}_{m,m-1}\tilde{R}^{\text{TM/TE}}_{m-1,m-2}e^{i2\beta_{m-1}(d_{m-1}-d_{m-2})}}.
\end{equation}
These are expressed in the usual Fresnel coefficients
\begin{eqnarray} 
  R^{\text{TM}}_{i,i-1} &=& \frac{\varepsilon_{i-1}\beta_i - \varepsilon_{i}
  \beta_{i-1}}{\varepsilon_{i-1}\beta_i + \varepsilon_{i}\beta_{i-1}}
\\
R^{\text{TE}}_{i,i-1} &=& \frac{\beta_i - \beta_{i-1}}{\beta_i + \beta_{i-1}}
\\
  T^{\text{TM}}_{i,i-1} &=& \frac{2
    \varepsilon_{i}\beta_{i-1}}{\varepsilon_{i-1}\beta_i 
  + \varepsilon_{i}\beta_{i-1}}
\\
  T^{\text{TE}}_{i,i-1} &=& \frac{2 \beta_i }{\beta_i + \beta_{i-1}}.
\end{eqnarray}

\section{Stationary phase method}
\label{sec:statphase}
To evaluate the two-dimensional integral in Eq.~(\ref{eqn:FTGreen}) 
we use the stationary phase method 
(see, e.g., Ref.~\cite{MandelWolf}).
There are two contributions to the phase of the integrand:
the Fourier transform term and the phase of the Green's function itself, which
depends on the phase of the Fresnel coefficients and a phase factor
that depends on the position $z'$ of the source.
Because the film thickness $d$ is much smaller than $\lambda$
the Fresnel coefficients vary slowly with $\bm{k}_p$. For this 
reason we need only to consider the position of the source. 
For brevity we here will only discuss 
the Green's function for the observation point $\bm{r}$ in region 1 
and the source position $\bm{r}'$ in region 3. Because of the general
form of the Green's function in this case (see Eqs.~(\ref{G13A}) - (\ref{G13B}))
integral (\ref{eqn:FTGreen}) can generally be written as
\begin{align}
    G_{\mu\nu}(\bm{r},\bm{r}',\omega)  &=    \int\frac{d^2\bm{k}_p}{(2\pi)^2}e^{i\bm{k}_p\cdot(\bm{r}-\bm{r}')}e^{-i\beta_1z}e^{i\beta_3|d-z'|}
\nonumber \\ &\times
  D_{\mu\nu}(k_x,k_y,\omega) \; ,
\end{align}
where the coefficients $ D_{\mu\nu}(k_x,k_y,\omega) $ are related to the Fresnel coefficients
and can be formally defined as
$ D_{\mu\nu}(k_x,k_y,\omega) = \fGF{\mu}{\nu} e^{i\beta_1z}e^{-i\beta_3|d-z'|}$.
The phase term of interest is then
\begin{equation}
\Phi={\bm{k}_p\cdot(\bm{r}-\bm{r}')}-\beta_1z+\beta_3|d-z'|.
\end{equation}
Recognizing that $z'>d$, and that $\bm{k}_p=(k_x, k_y, 0)$ is the same
 in region 1 and 3 we get
\begin{equation}
\Phi=k_x(x-x')+k_y(y-y')-\beta_1z+\beta_3(z'-d)
\end{equation}
The phase is then expanded in powers of $k_1r$. To do so we introduce the quantities
\begin{eqnarray}
  s_x  &=&  \frac{x}{r} \quad  , \quad  s_y = \frac{y}{r} 
\\
  s_x'  &=&  \frac{x'}{r}  \quad  , \quad s_y' = \frac{y'}{r} 
\\
  p  &=&  \frac{k_x}{k_1}  \quad  , \quad q = \frac{k_y}{k_1}
\\
  s_z &=& \frac{z}{r}  = (1 - s_x^2 - s_y^2)^{1/2} 
\\
  s_z'&=& \frac{z'}{r}  =  (1 - s_x'^2 - s_y'^2)^{1/2} 
\\
  m&=& -\frac{\beta_1}{k_1}  = -(1 - p^2 - q^2)^{1/2} 
\\
  m'&=& \frac{\beta_3}{k_1}  =  +\left(\left(\frac{k_3}{k_1}\right)^2 - p^2 - q^2\right)^{1/2} \; ,
\end{eqnarray}
where $(s_x, s_y, s_z)$ determine the direction of the observation point relative to the source,
and $p, q$ are re-scaled integration variables.
We can then express the phase as
\begin{equation}
\Phi = k_1r \left(p(s_x-s_x')+q(s_y-s_y')+ms_z+m's_z'\right).
\end{equation}

The integral is evaluated in the limit of large $k_1r$. The stationary points are determined
by setting the derivatives $\partial \Phi / \partial p $ and $\partial \Phi / \partial q $  to zero
which yields
\begin{eqnarray}
\frac{s_x-s_x'}{s_z+s_z'\frac{m}{m'}}=\frac{p}{m} & \textrm{and} & \frac{s_y-s_y'}{s_z+s_z'\frac{m}{m'}}=\frac{q}{m}.
\end{eqnarray}

Because $r\gg r'$ we generally have $s_i' \ll s_i$. This is not true if we observe the far field very
close to the $z$-axis (so that $s_x\approx s_x'$) or very close to the plane of the interface (so that $s_z\approx s_z'$). Ignoring these special cases we can simplify the stationary points to
\begin{eqnarray}
\frac{s_x}{s_z}=\frac{p}{m} & \textrm{and} & \frac{s_y}{s_z}=\frac{q}{m},
\end{eqnarray}
which is the usual result of the far field approximation that $\bm{k}$ and $\bm{r}$ are parallel.

To determine the contribution of the critical points to the integral the second order partial derivatives at the stationary points must be evaluated. These evaluated derivatives are

\begin{align}
  \frac{ \partial^2 \Phi}{\partial x^2} 
  &=  
   -\left(1+\left(\frac{s_x}{s_z}\right)^2\right),
\\
  \frac{ \partial^2 \Phi}{\partial y^2} 
  &= 
  -\left(1+\left(\frac{s_y}{s_z}\right)^2\right),
\\
  \frac{ \partial^2 \Phi}{\partial x \partial y} 
   &=
  -\frac{s_xs_y}{s_z^2}.
\end{align}
If we parametrize the stationary point as
$p=\frac{k_x}{k_1}=\sin{\phi}\cos{\theta}$, and $q=\frac{k_y}{k_1}=\sin{\phi}\sin{\theta}$,
the Green's function in stationary phase approximation becomes
\begin{align}
 G_{\mu\nu}(\bm{r},\bm{r}',\omega) &= 
  \frac{i}{2\pi}\left(\frac{k_i}{r}\frac{z}{r}\right)e^{ik_1r}e^{-ik_1\sin{\phi}(\cos{\theta}x'+\sin{\theta}y')}
\nonumber \\
  &\times e^{-i\beta_1z}e^{i\beta_3|d-z'|}
\nonumber \\ 
  &\times {D}_{\mu\nu}(k_1\sin{\phi}\cos{\theta},k_1\sin{\phi}\sin{\theta},\omega).
\end{align}
The terms $\beta_j=k_{z,i}$ are evaluated at
$k_p=k_1\sin{\phi}$, such that
$\beta_j=\sqrt{k_j^2-k_1^2\sin^2{\phi}}$. 
The effect of the stationary phase method on the integral is a leading
term that is proportional to $z$. This results in the vanishing of
terms near the boundary with the exception of those tied to $TE$
radiation. 
Expressing $ {D}_{\mu\nu}$ in terms of $\tilde{G}_{\mu\nu}$ then yields Eq.~(\ref{eqn:statphaseGxx}).

\end{document}